\documentclass[sigconf, nonacm, screen]{acmart}
\usepackage[inline]{enumitem}
\usepackage[dvipsnames]{xcolor}
\usepackage{todonotes}
\usepackage{lipsum}
\usepackage{subcaption}
\usepackage{cleveref}
\usepackage{makecell}
\usepackage{graphicx}
\usepackage{tabularx}
\usepackage{bm}
\usepackage{pifont}

\newcommand{\sparagraph}[1]{\vspace{1mm}\noindent {\bf #1}}
\newcommand{\thesystem}{JetStream}

\usepackage{adjustbox}
\usepackage{array}
\newcolumntype{R}[2]{%
    >{\adjustbox{angle=#1,lap=\width-(#2)}\bgroup}%
    l%
    <{\egroup}%
}
\newcommand*\rot[2]{\multicolumn{1}{R{#1}{#2}}}%

\newcommand{\Sup}{\textcolor{ForestGreen}{\textbf{\ding{51}}}}
\newcommand{\NSup}{\textcolor{red}{\textbf{\ding{55}}}}
\newcolumntype{Y}{>{\centering\arraybackslash}X}

\begin{document}

\title{JetStream: Generating Query Accelerators for Existing Database Systems}

\author{Akhilesh Balasingam}
\orcid{0009-0003-8637-896X}
\affiliation{%
  \institution{MIT CSAIL}
  \city{Cambridge}
  \state{MA}
  \country{USA}
}
\email{akhilvb@mit.edu}

\author{Amadou Ngom}
\orcid{0009-0000-8422-5212}
\affiliation{%
  \institution{MIT CSAIL}
  \city{Cambridge}
  \state{MA}
  \country{USA}
}
\email{ngom@mit.edu}

\author{Geoffrey X. Yu}
\orcid{0009-0005-3186-1465}
\affiliation{%
  \institution{MIT CSAIL}
  \city{Cambridge}
  \state{MA}
  \country{USA}
}
\email{geoffxy@mit.edu}

\author{Tim Kraska}
\orcid{0009-0003-2414-2759}
\affiliation{%
  \institution{MIT CSAIL}
  \city{Cambridge}
  \state{MA}
  \country{USA}
}
\email{kraska@mit.edu}

\begin{abstract}
% TODO: clunky
Recent work has shown that LLMs can synthesize highly specialized database systems for fixed workloads, but existing approaches typically assume static data and replace the database's native storage with generated representations. We present \thesystem{}, a system for generating query-specific accelerators that instead extend an existing DBMS. 

\thesystem{} consists of three parts. First, a staged, measurement-driven agentic workflow generates and optimizes query-specific accelerators, including persistent auxiliary state when beneficial. Second, a fixed, engine-neutral substrate provides the common interfaces for execution, transaction coordination, state management, and maintenance. Third, a separate synthesis workflow generates engine-specific backend adapters that connect the substrate to the underlying engine. For stateful accelerators, \thesystem{} also generates maintenance logic and uses a runtime policy to choose among incremental maintenance, rebuilds, and lazy repair as the database changes.

On TPC-H at SF=20, stateful accelerators generated by \thesystem{} achieve an $833\times$ geomean read-only speedup over DuckDB, compared with $34.07\times$ for GenDB and $12.35\times$ for Bespoke OLAP. Under TPC-H refreshes every 60 seconds, \thesystem{} maintains a $375\times$ geomean workload speedup. We also show that \thesystem{} generalizes to new, unseen workloads, achieving geomean read-only speedups of $102\times$ over DuckDB and $486\times$ over PostgreSQL. These results show that aggressive generated specialization can be integrated seamlessly with existing DBMSes, and support dynamic workloads.
\end{abstract}

\maketitle

\section{Introduction}\label{sec:intro}

% \singlefigplaceholder{2.5in}{Motivating result: Workload speedup on our system with
% comparison to GenDB, BespokeOLAP, Base database system (probably DuckDB). Construct
% a workload with sufficient writes. Plot speedup, auxiliary
% space usage. Result should show Ours > \{GenDB, BespokeOLAP\} > Base system. It
% might also be that Ours > Base > \{GenDB, BespokeOLAP\} with sufficient writes.
% We should include both this realistic setting (i.e., with writes) and a
% read-only setting (the scenario GenDB/BespokeOLAP were designed for).}

% TODO:
% move to bar chart, move mem footprint to later section
% add BRAD in 1 sentence

\begin{figure}[t]
    \centering
    \includegraphics[width=\columnwidth]{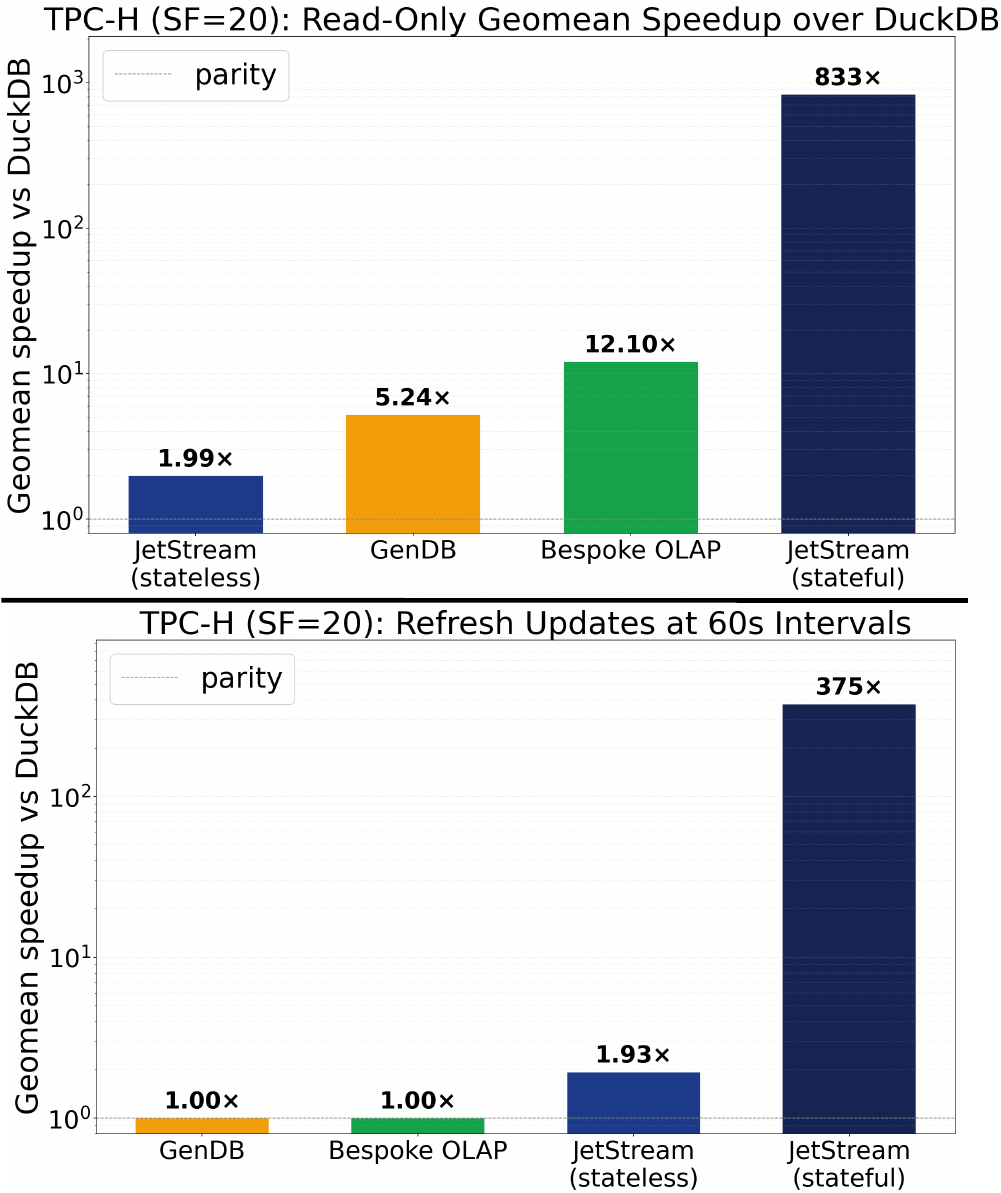}
    \caption{\textbf{\thesystem{} combines aggressive read acceleration with
    support for realistic write workloads.} On TPC-H SF=20, stateful
    \thesystem{} achieves substantially higher read-only speedup than prior
    generated systems (top). Under
    TPC-H refreshes every 60 seconds, \thesystem{} retains a $375\times$
    geomean workload speedup over DuckDB, while systems without write
    maintenance fall back to parity with the base DBMS (bottom).}
    \label{fig:headline}
    \vspace{-1em}
\end{figure}

\newcommand{\Yes}[1]{%
  \makecell[l]{\makebox[\linewidth][c]{\large\Sup{}}\\[0.1em] #1}}
\newcommand{\No}{\makecell[c]{\large\NSup{}}}

\begin{table}[t]
\caption{No existing query accelerator synthesis system supports all desired practical capabilities.}\label{tbl:qual-comparison}
\vspace{-0.5em}
\small
\begin{tabularx}{\columnwidth}{@{}ll YYYY@{}}
  \toprule
  \textbf{System} & &
    \rot{90}{1em}{\makecell[l]{\textbf{Accelerates}\\\textbf{known queries}}} &
    \rot{90}{1em}{\makecell[l]{\textbf{Supports}\\\textbf{ad-hoc queries}}} &
    \rot{90}{1em}{\makecell[l]{\textbf{Preserves accel.}\\\textbf{under writes}}} &
    \rot{90}{1em}{\makecell[l]{\textbf{Uses existing}\\\textbf{DBMS storage}}} \\
  \midrule
  \addlinespace[0.75em]
  \makecell[l]{Castor~\cite{castor-feser20}} & &
    \Yes{Deductive\\program\\synthesis} &
    \No{} & \No{} & \No{} \\
  \addlinespace[0.75em]
  \makecell[l]{GenDB~\cite{gendb}} & &
    \Yes{Agentic\\synthesis\\workflow} &
    \No{} & \No{} & \No{} \\
  \addlinespace[0.75em]
  \makecell[l]{Bespoke\\OLAP~\cite{bespoke-olap}} & &
    \Yes{Agentic\\synthesis\\workflow} &
    \Yes{Separate\\DBMS\\fallback} &
    \No{} & \No \\
  \midrule
  \addlinespace[0.75em]
  \makecell[l]{\textbf{\thesystem{}}\\\textbf{(this work)}}  & &
    \Yes{Agentic\\synthesis\\workflow} &
    \Yes{Underlying\\DBMS\\execution} &
    \Yes{Intelligent\\aux. state\\maintenance} &
    \Yes{Using a\\general\\substrate} \\
  \bottomrule
\end{tabularx}
\end{table}

% Story:
% - Relational DBMSes designed to be general
% - If you know which queries you expect to run over and over, you can clearly
%   run it faster with a specialized system
% - Up until now, you would only build such a system if the workload is
%   particularly important (e.g., revenue generating)
% - Now with LLMs, this calculation has changed
%
% - Recent work has explored using LLMs to synthesize query engines for specific
%   query templates
% - Describe GenDB, BespokeOLAP
% - Both systems good for static databases
% - However, we argue that both systems lack capabilities needed for practicality
% - Supporting unseen queries, preserving perf. benefits under writes, and not duplicating the storage
% - Achieving the properties above is challenging because [...]

% - Introduce JetStream, first by highlighting the performance results
% - Then point back to Table 1 as we go
% - Describe correctness guarantees -> same as GenDB, Bespoke OLAP - empirical
%   testing, which we argue is no worse than existing systems that also rely on
%   testing

% - Give summary of key results, contributions list
% This generality, however, leaves performance on the table.

Relational database systems are designed to be general-purpose: they support
diverse schemas and data and can execute any query expressible in their SQL
dialect.
But this generality comes at the cost of performance.
%
% Database systems rely on general-purpose query processing algorithms (e.g., hash
% joins), data layouts (e.g., sorted columnar), and data structures (e.g., ordered
% indexes), rather than implementations specially constructed for a particular
% query template and dataset.
%
For example, while an aggregation can be generally implemented using a hash
table, if the set of groups are known ahead of time, we could instead use a
direct-mapped array and eliminate the hash-probing overhead altogether.

This example shows how knowledge of a query workload and its data can enable
specialized execution strategies, which we call \emph{accelerators}, that
outperform general-purpose database systems.
Historically, building such accelerators required substantial engineering
effort, and by design, sacrificed generality.
% by specializing to a narrow workkload.
%
Such systems were thus typically built only for query workloads important enough
to justify both the engineering effort and loss of generality (e.g.,
performance- or revenue-critical applications~\cite{scuba-abraham13,
napa-agiwal21, f1lightning-yang20}).
But recent advances in large language model (LLM) code
generation~\cite{openai2025codex, anthropic2025claudecode, wang2024openhands}
are changing this trade-off by making it practical to automatically
construct specialized query accelerators for known workloads~\cite{gendb,
gptdb-trummer23, bespoke-olap}. This raises a natural research question: can we
obtain the performance benefits of specialization without giving up the
generality and practicality of a conventional database system?

Systems such as GenDB~\cite{gendb} and Bespoke OLAP~\cite{bespoke-olap}
take a step in this direction, both synthesizing query engines for a known
workload by specializing its storage layout, data structures, and query
implementations.
However, as outlined in \Cref{tbl:qual-comparison}, both systems make design
decisions that limit their practicality.
First, they rewrite the base data itself, which complicates writes, potentially
requires data to be rewritten if the workload changes, and forces users to
migrate their data into the specialized format.
Second, their specialized data structures assume static data and lack a
mechanism for incremental maintenance. After writes, they either return stale
results or forgo their performance benefits altogether.
Third, GenDB in particular sacrifices generality entirely by not supporting
ad-hoc queries.
We argue that a practical system leveraging synthesized query accelerators
should retain the generality of a conventional DBMS, supporting writes and
ad-hoc queries using its native storage, while delivering accelerated
performance for known, pre-declared query templates.

% Challenges
% - Reward hacking
% - Reduce token bloat (reduce costs)
%   - Negative memory
% - Common substrate
Two challenges make achieving this difficult in practice. First, accelerator integration is susceptible to reward hacking \cite{metr-2025-recent-reward-hacking}: modern coding agents can modify the system and invoke their own tests and benchmarks, making it unsafe to let the same agent optimize an accelerator and judge whether it is correct or faster. Apparent improvements can come from weakening correctness checks or exploiting measurement artifacts rather than actually improving the system, so correctness and performance evaluation must be owned by an external mechanism with fixed acceptance criteria. Second, integrating accelerators across multiple DBMSes requires more than generating query code, because engines expose different scan APIs, mutation paths, and transaction mechanisms. Encoding these details directly into every accelerator would tightly couple accelerator logic to a particular engine and force the synthesis process to repeatedly rediscover the same integration logic.

We present \thesystem{}: a new practical system that synthesizes query-specific
accelerators and incorporates them \emph{within} an existing database system.
\thesystem{} addresses the first challenge with a measurement-driven synthesis workflow in which an external orchestrator, rather than a coding agent, owns correctness and performance evaluation. It addresses the second with a fixed, engine-neutral substrate that separates query-specific accelerator logic from engine-specific database integration.
This separation allows \thesystem{} to specialize aggressively without replacing the underlying database. Accelerators can execute directly over native base tables or maintain query-specific auxiliary state, while the DBMS remains responsible for canonical storage, transactions, recovery, and concurrency. Queries without accelerators continue to execute normally, and accelerators can be added, removed, or regenerated without changing the representation of the base data.
Like existing systems~\cite{gendb, bespoke-olap}, \thesystem{} applies
differential testing during accelerator synthesis and uses a final audit stage
(\Cref{sec:correctness}) to empirically validate that its accelerators match the
semantics of their corresponding SQL queries.
\thesystem{} builds on top of our prior work on Tailwind~\cite{tailwind-yu26},
which non-invasively integrates manually-written accelerators into existing
database systems; \thesystem{} is the first step towards automatically
generating such accelerators.

As shown in Figure \ref{fig:headline}, our evaluation shows that this design provides both aggressive specialization and practical update support. On TPC-H at SF=20, stateful \thesystem{} accelerators achieve an $833\times$ geomean read-only speedup over DuckDB, compared with $34.07\times$ for GenDB \cite{gendb} and $12.35\times$ for Bespoke OLAP \cite{bespoke-olap}. With TPC-H refreshes every 60 seconds, \thesystem{} retains a $375\times$ geomean workload speedup. On SEC-EDGAR, a workload chosen to reduce potential LLM training-data overlap, \thesystem{} shows that it can generalize well, achieving $102\times$ and $486\times$ geomean read-only speedups over DuckDB and PostgreSQL, respectively. 

In summary, we make the following contributions: 
\begin{itemize}[leftmargin=*] 
\item \textbf{A measurement-driven agentic workflow for database accelerator generation.} \thesystem{} separates accelerator analysis, implementation, optimization, and validation; grounds optimization in profiling and controlled measurements; and safely synthesizes multiple query accelerators concurrently. 
\item \textbf{Generated accelerators that extend, rather than replace, an existing DBMS.} We introduce an engine-independent substrate together with generated backend adapters that allow specialized query code to execute directly over native database storage. The same framework supports stateless or stateful accelerators and single- or multi-threaded execution. 
\item \textbf{A write-aware framework for generated auxiliary state.} Stateful accelerators include query-specific maintenance logic that incrementally updates, rebuilds, or invalidates auxiliary state based on the expected cost of writes and future queries, together with a transactional protocol that keeps generated state consistent with committed base data after updates. 
\end{itemize}

\section{System Overview \& Details}

\begin{figure*}[t]
    \centering
    \includegraphics[width=\textwidth]{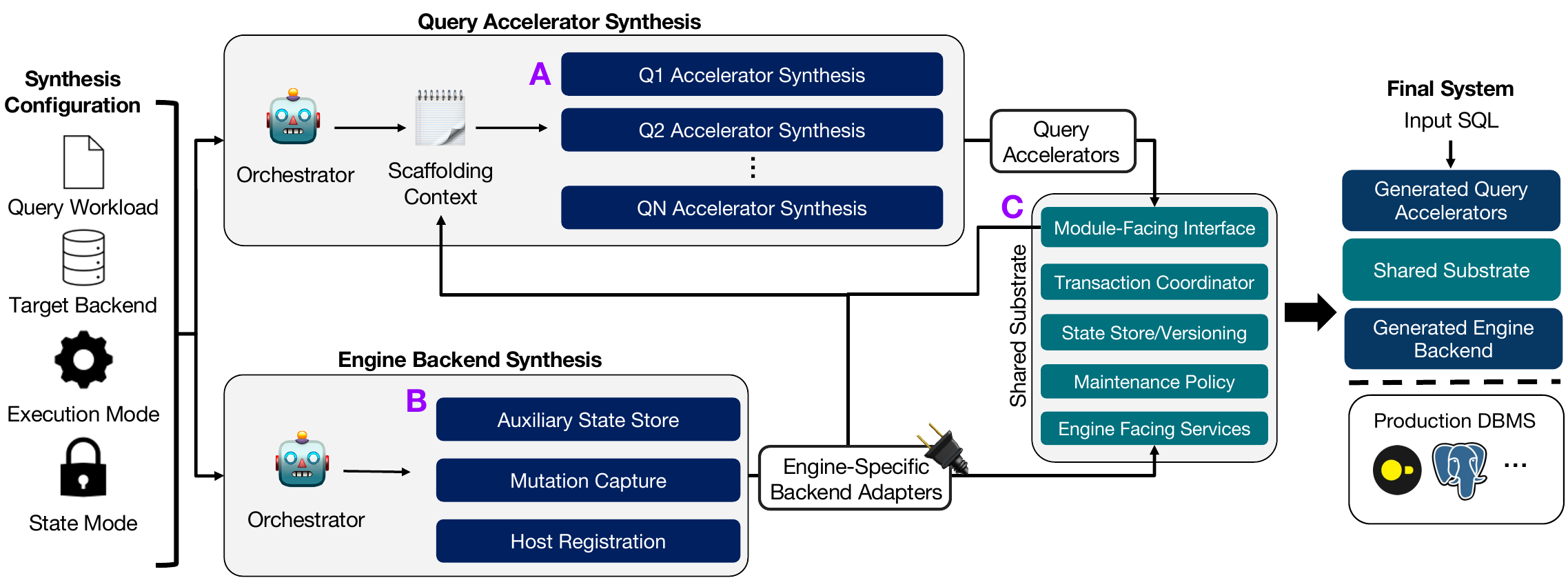}
    \caption{\thesystem{} system overview. 
    \textbf{(A)} Concurrent synthesis of query-specific accelerators
    (Section~\ref{sec:agentic-workflow});
    \textbf{(B)} generation of engine-specific backend adapters
    (Section~\ref{sec:system-architecture}); and
    \textbf{(C)} the fixed shared substrate connecting generated accelerators
    to the underlying DBMS
    (Section~\ref{sec:system-architecture}).}
    \label{fig:system}
\end{figure*}

% \begin{itemize}[leftmargin=*]
%   \item Discuss agentic workflow
%   \item Discuss system architecture
%   \item Describe key idea for keeping auxiliary state up-to-date
%   \item Describe correctness assumptions
%   \item Probably: more detailed discussion about handling writes to the
%     auxiliary state
% \end{itemize}

% \noindent
% \gy{The system overview and details section will likely change. We should
% probably just start by writing out everything that we deem interesting and then
% we can work on distilling/reorganizing it into sections/subsections for the
% paper.}

\thesystem{} takes as input a query workload together with a synthesis configuration specifying the target database engine, whether execution is single- or multi-threaded, and whether persistent auxiliary state is permitted. As shown in Figure~\ref{fig:system}, \thesystem{} generates query-specific accelerators for the workload and the engine-specific backend needed to deploy them on an existing DBMS. Accelerators can execute directly over native DBMS storage or additionally maintain query-specific auxiliary structures.

The generated accelerators and backend compose through a fixed, engine-neutral substrate, while the underlying DBMS remains responsible for canonical storage and transaction processing. The remainder of this section follows the three components highlighted in Figure~\ref{fig:system}.
% We first describe how \thesystem{} generates and optimizes query accelerators, including concurrent synthesis across a workload consisting of many queries. We then present the shared substrate and generated engine backends, before describing how stateful accelerators 
\subsection{Agentic Workflow}
\label{sec:agentic-workflow}
% - user specifies synthesis envelope
% - staged pipeline (what each stage does)
% - evidence driven candidate selection
% - global negative memory
% - automatic experiment synthesis (dynamic profiling) 
% - rigorous testing defined at implementation time, expanded at audit to prevent cheating
% - to avoid hillclimbing for benchmarking in attempts we do sampled write freq so it gets a diversity of signal on write maintenance path
% - evidence compression btwn attempts

\thesystem{} generates each query accelerator through a staged, \\
measurement-driven workflow. Although modern coding agents can inspect a repository, modify code, and invoke tests and benchmarks within a single session, accelerator synthesis is fundamentally a multi-stage design problem with delayed and noisy feedback. Early decisions about execution strategy and, when applicable, auxiliary state organization determine the space of implementations that can be explored, but their consequences only become observable after a concrete system is built and exercised under load. Conflating these phases into a single unconstrained agent loop risks premature commitment to suboptimal designs and makes it difficult to attributed observed performance to specific design choices. Moreover, if the agent is responsible for both modifying the system and evaluating its correctness and performance, the evaluation signal becomes part of the agent's own decisions, weakening the reliability of acceptance criteria iterations. Empirically, we found that even strong coding agents, including Codex with GPT-5.6.-Sol, did not reliably synthesize a complete set of query accelerators from only a workload description and a target database engine.

\begin{figure}[t]
    \centering
    \includegraphics[width=\columnwidth]{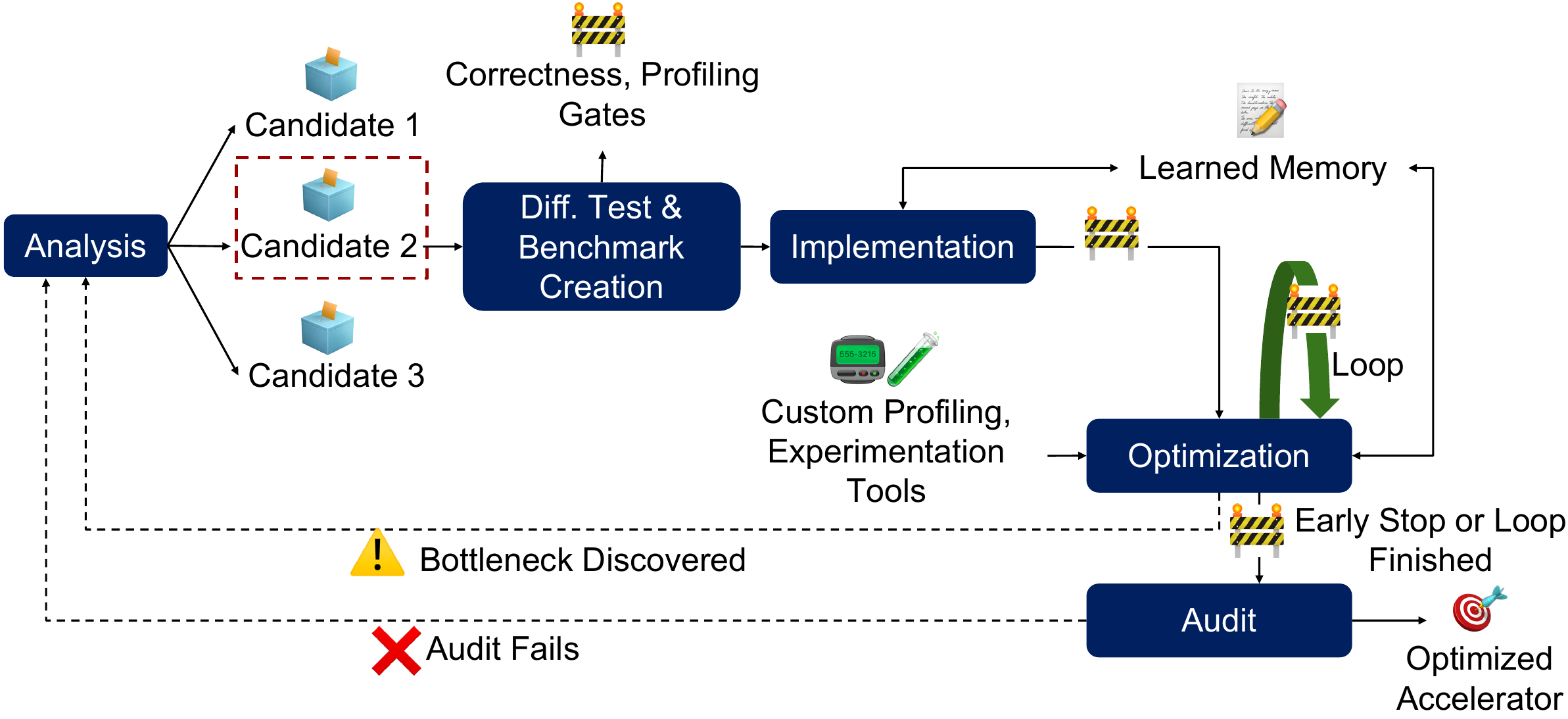}
    \caption{\thesystem{}'s query-accelerator synthesis workflow.}
    \label{fig:agent-synthesis}
\end{figure}

\thesystem{} instead structures accelerator generation as the workflow shown in Figure \ref{fig:agent-synthesis}. The pipeline separates high-level design, implementation, optimization, and audit, while an external orchestrator owns correctness and performance evaluation.

\subsubsection{Staged Accelerator Synthesis}

Each stage in Figure \ref{fig:agent-synthesis} is executed by a separate instance with stage-specific context, instructions, and tools. The analysis agent receives the query, synthesis configuration, and repository context needed to understand the available substrate and existing accelerator implementations. Later agents receive the artifacts produced by earlier stages, together with the tools relevant to their task, such as build, test, benchmark, and profiling interfaces. This keeps each agent focused on a narrow objective while preserving the information needed across stages.

The analysis stage proposes several materially different accelerator designs. These candidates specify the high-level execution strategy, and, for stateful accelerators, the organization of auxiliary state and its maintenance strategy. \thesystem{} selects among these alternatives before committing to an implementation.

The implementation agent realizes the selected design against the shared substrate. Before optimization begins, \thesystem{} also constructs the differential tests and benchmarks used to evaluate the accelerator, preventing later optimization from weakening correctness checks or changing the benchmark to favor a candidate. After optimization converges, a separate audit agent expands validation beyond the tests visible during implementation and optimization. An audit failure returns the pipeline to analysis; otherwise, the accelerator is finalized.

\subsubsection{Evidence-Guided Optimization}

Optimization proceeds through measured candidate attempts rather than free-form code refinement. For each attempt, the orchestrator builds the candidate, runs the required correctness gates, and compares it against the best validated implementation using controlled A/B benchmarks. A candidate becomes the new best only if it remains correct and improves the optimization objective.

When aggregate latency does not reveal the bottleneck, the optimization agent can synthesize additional instrumentation and experiments, such as phase timers and query-specific counters. These measurements help distinguish costs from base scans, auxiliary-state lookup, decoding, synchronization, or maintenance. If the evidence shows that the current state organization or execution strategy has reached a structural limit, the pipeline returns to analysis rather than continuing to accumulate local optimizations.

For stateful accelerators, the optimization signal includes both reads and writes. Rather than evaluating every attempt at a single write frequency, \thesystem{} samples multiple write frequencies across attempts, exposing the optimizer to different read-versus-maintenance tradeoffs. 

The pipeline also carries evidence forward across attempts. Failed hypotheses, performance regressions, and identified bottlenecks are retained as negative experience, while successful changes and useful profiles are preserved. This evidence is compressed between attempts so later agents avoid repeating failed work without carrying the full conversational history.

\subsubsection{Concurrent Workload Synthesis}
% - worker private state
% - explicit shared state
% - ownership-aware recovery
% - token-cost cutting measures like repo context preloaded, etc.

\thesystem{} synthesizes multiple query accelerators concurrently rather than processing a workload serially. Each query runs as an asynchronous worker that advances independently through analysis, implementation, and optimization, so a slow query does not impose a stage barrier on the rest of the workload.

Naively running multiple coding agents against one repository is unsafe because query implementations ultimately compile into a shared library and touch common registration and build surfaces. \thesystem{} therefore gives each worker a private source worktree, build directory, and agent runtime. Query-owned files remain private, shared registration files are coordinated through an explicit lease, and workers cannot modify another query's generated implementation.

The orchestrator also distinguishes failures in a worker's own code from failures caused by shared or peer state. In the latter case, it restores the affected files from the most recent validated snapshot rather than spending agent turns debugging foreign code. Workers additionally receive a compact shared repository context describing stable integration points and existing accelerators, reducing repeated repository discovery while keeping the individual synthesis trajectories isolated.

\subsection{System Architecture}
\label{sec:system-architecture}
% - engine-agnostic substrate
% - engine adapter generation
% - stateless over native DBMS storage, stateful over new interface

\thesystem{} separates query-specific accelerator logic from engine-specific database integration through a fixed shared substrate. Generated accelerators do not call DBMS APIs directly; instead, they interact only with engine-neutral substrate interfaces, while generated backend adapters translate those interfaces to the native mechanisms of the target database. This yields a three-layer runtime structure: query accelerators define \emph{what} computation and auxiliary state to use, the substrate provides common transaction coordination and state-management mechanisms, and the backend adapter implements the engine-specific I/O, storage, and transaction hooks. 

At runtime, each accelerator is exposed through the host DBMS using its normal extension mechanism---for example, as a table function in DuckDB or an extension function in PostgreSQL. Invoking this entry point dispatches into the generated query module, which executes through the shared substrate and reaches the underlying engine only through the generated backend adapter.

\subsubsection{Shared Accelerator Substrate}
\begin{figure}[t] \centering \includegraphics[width=0.6\columnwidth]{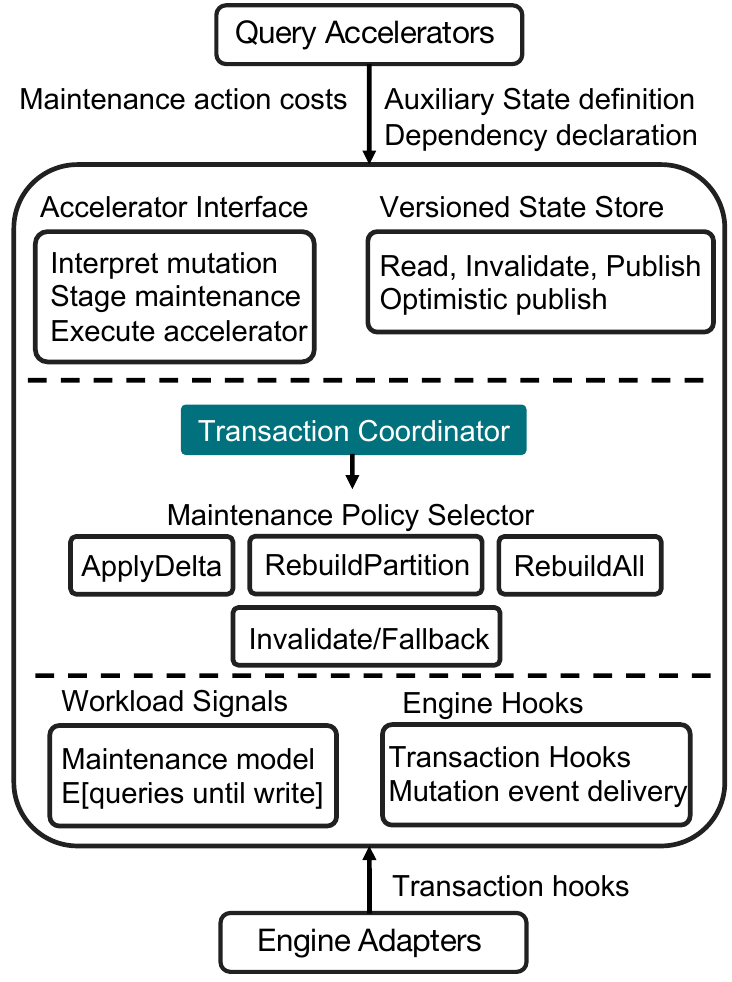} 
\caption{The fixed shared substrate separates query-specific accelerator logic from engine-specific database integration.} 
\label{fig:shared-substrate} \end{figure}

Figure~\ref{fig:shared-substrate} details the fixed substrate and the interfaces supplied by generated accelerators and engine backends. Its transaction coordinator registers accelerators and routes mutations according to their declared dependencies; its state layer tracks the validity and versions of its auxiliary state; and its maintenance layer selects how state should be repaired after writes. The substrate additionally exposes portable native-scan requests that the generated backend translates into engine-native table scans.

The same substrate supports both stateless and stateful accelerators. Stateless accelerators request the columns and predicates they need through an engine-neutral scan interface; the backend maps that request onto the DBMS's native storage scan and streams the resulting data to the generated accelerator. Stateful accelerators additionally store persistent query-specific structures through the substrate's state interface. These structures are organized into \emph{partitions}, where a partition is a query-defined unit of auxiliary state, such as a month, date range, or key range. This partitioning is independent of any physical partitioning used by the underlying DBMS and allows maintenance to target only the affected portion of an accelerator's state. This boundary keeps query-specific state and execution logic independent of engine-specific storage and transaction mechanisms.

% The substrate also isolates accelerator semantics from engine semantics. Query modules determine how a query should be executed, what auxiliary state should be built, and how a mutation affects that state. They do not interpret engine transaction callbacks or manage shared memory. Conversely, the substrate and backend do not encode query-specific logic. This boundary allows the same generated accelerator to execute against different database engines.

\subsubsection{Generated Engine Backends}

For each target DBMS, \thesystem{} generates a backend that implements the engine-facing side of the substrate. The backend provides four core pieces of functionality: native base-table scans, capture and translation of inserts, updates, and deletes, transaction lifecycle callbacks, and storage for auxiliary-state images. It also provides the extension or table-function glue needed to expose generated accelerators through the host database.

The native scan path stays inside the DBMS. In DuckDB, the backend binds scan requests to the engine's native table-storage APIs, including available filter pushdown, and streams data directly to the generated accelerator. PostgreSQL uses its table access methods to perform the corresponding scans before passing the results through the same engine-neutral representation. Thus, state construction operate over the canonical base tables without reissuing the query through a client SQL interface.

The auxiliary-state representation adapts to the execution model of the underlying engine. DuckDB runs in process with the accelerator runtime, so the backend can retain typed in-memory state without serialization on the query hot path. PostgreSQL uses separate backend processes and shared memory, so the backend serializes auxiliary state at the process boundary and reconstructs the same logical query state before the accelerator consumes it. 

Finally, the backend translates native write and transaction events into the common substrate protocol. Inserts, updates, and deletes become engine-neutral mutations, while pre-commit, commit, abort, and savepoint events drive the corresponding maintenance lifecycle. The adapter is the only layer that knows how these events are exposed by a particular engine. For DuckDB, which does not currently expose all transaction-update hooks required by our maintenance protocol, we add a small set of hooks to its transaction-handling internals to surface these events to the generated backend.

\subsection{Auxiliary State Maintenance}
\label{sec:maintenance}
% - high-level maintenance policy - writes trigger one of several actions (no op, apply delta, rebuild partition, rebuild all, invalidate and fallback), relevance filtering at substrate level, miantenance action selected based on estimated immediate, future query, interference costs - read speed + write/maintenance cost jointly optimized

\begin{figure}[t] \centering \includegraphics[width=\columnwidth]{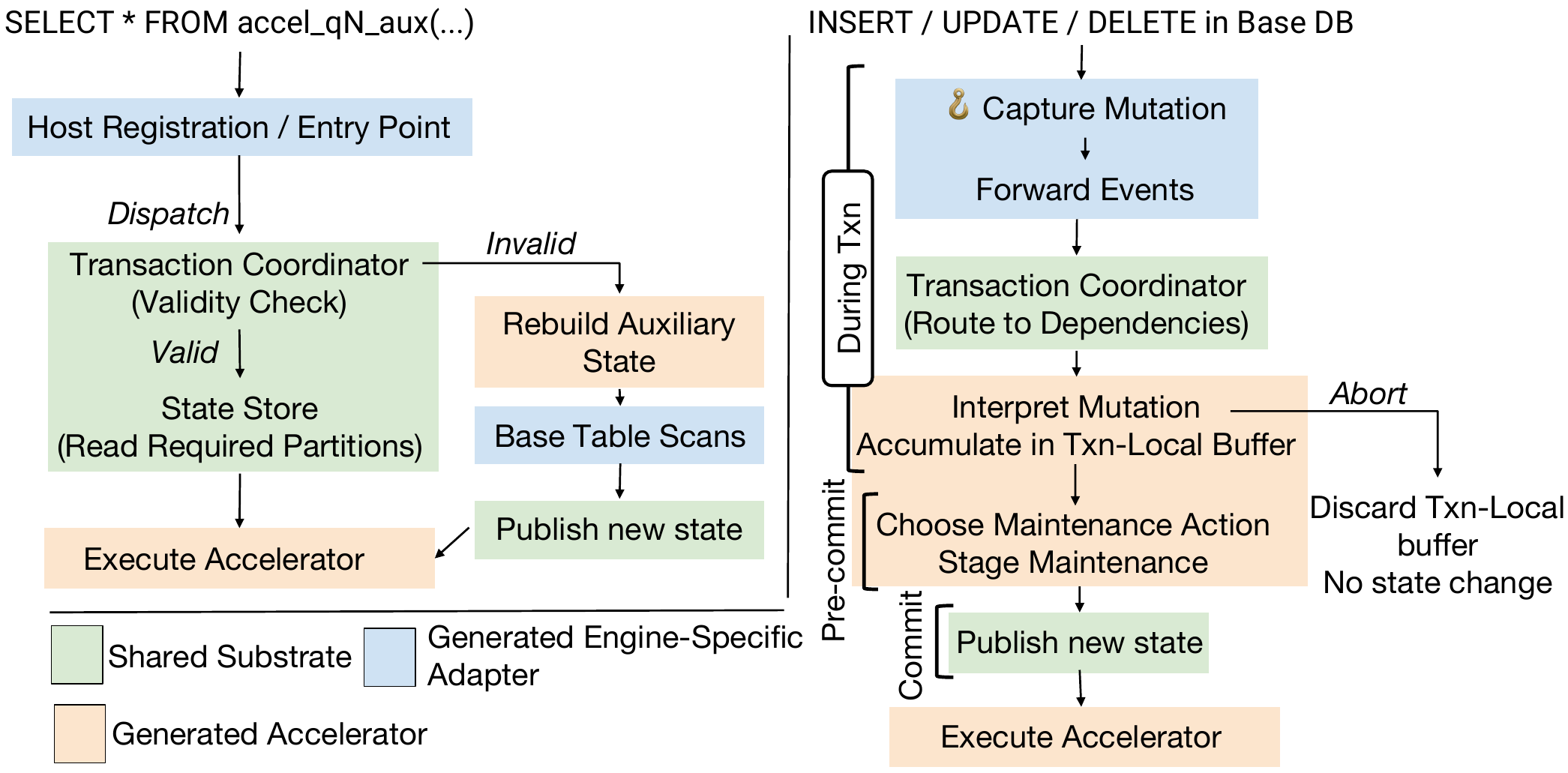} 
\caption{Read and write paths for a stateful accelerator. 
% Reads use valid auxiliary state when available and otherwise fall back to execution over the canonical base tables. Writes are captured by the generated backend, routed through the shared substrate, interpreted by the query accelerator, and staged for maintenance before publication at commit.
} \label{fig:read-write-path} 
\end{figure}

Stateful accelerators must keep their auxiliary structures synchronized with the canonical base tables without making every write pay the cost of fully rebuilding state. \thesystem{} therefore separates maintenance into two parts: a policy that chooses how an accelerator should respond to a transaction's writes, and a transactional protocol that stages and publishes the resulting state changes. Figure~\ref{fig:read-write-path} summarizes where these decisions occur in the read and write paths.

\subsubsection{Maintenance Policy}

When a transaction modifies the database, \thesystem{} first filters the write using each accelerator's declared dependencies and query-specific interpretation of the mutation. Writes that cannot affect an accelerator require no maintenance. Relevant writes are accumulated over the transaction together with the number of rows touched and the set of auxiliary-state partitions affected. 

At pre-commit, each accelerator chooses one maintenance strategy for the transaction. It can incrementally update the affected state, rebuild only the affected partitions, rebuild the full auxiliary structure, or invalidate the affected state and defer to the read path (where a subsequent query can lazily publish a repaired state). The available choices depend on the generated accelerator: for example, incremental maintenance is used only when the accelerator provides a valid update path, while partition rebuilding requires that the affected state can be localized.

To choose among the valid actions, the substrate maintains a runtime cost model for each accelerator. The model combines measurements of query and maintenance costs with properties of the current transaction and workload behavior. Thus, the decision can change as writes become larger or more frequent rather than relying on a fixed maintenance policy. For an action $a$,
\[ 
C(a) = C_{\mathrm{immediate}}(a) + C_{\mathrm{future}}(a), 
\]
where the first term captures work paid by the current write and the second captures the expected read-side cost before the next write. 

In our implementation, these estimates use simple measured models:
\[ 
C_{\mathrm{delta}} = c_0 + c_r R, \qquad C_{\mathrm{part}} = P c_p, \qquad C_{\mathrm{full}} = c_f, 
\]
while invalidation trades low immediate cost against future read overhead:
\[ 
C_{\mathrm{invalidate}} = c_i + Q\left(T_{\mathrm{fallback}} - T_{\mathrm{hit}}\right). 
\]

Here, both $R$, the number of rows touched, and $P$, the number of affected auxiliary-state partitions, come from the current transaction. The maintenance-cost parameters and query lantencies are calibrated from observed executions, and $Q$ is estimated from the observed number of queries between writes. The lowest-cost eligible action is selected. This simple model captures the basics of the central tradeoff: incremental maintenance is attractive for small writes, localized rebuilds for small affected regions, while invalidation is cheap immediately but gives up future accelerated reads.

% The maintenance decision is made once per accelerator and writer transaction, but the resulting updates can be applied at partition granularity. Incremental maintenance therefore replaces only the affected immutable state images and leaves unrelated partitions unchanged.

\subsubsection{Transactional Maintenance Protocol}

The the transactional protocol determines \emph{when} changes to the auxiliary state become visible. When the DBMS observes an insert, update, or delete, the generated backend converts the engine-specific event into a common mutation representation containing the affected table, operation type, old and new row values, and changed columns. The shared substrate routes the mutation only to accelerators that depend on the modified table, as shown in Figure \ref{fig:read-write-path}.

Each accelerator then interprets the mutation according to its own state organization, identifies the affected partitions, and accumulates query-specific changes in transaction-local state. These staged changes remain invisible to other transactions. At pre-commit, the accelerator applies the maintenance decision described above to the accumulated transaction-local changes. For incremental maintenance, it materializes replacement state for the affected partitions; for a rebuild, it records it to lazily rebuild on the next missed read.

At commit, \thesystem{} installs the resulting state. Rebuilds rescan the canonical base tables, reconstruct the required partitions or full structure, invalidate the corresponding old state, and publish the replacement. Incremental updates are published optimistically: each new partition image is associated with the version from which it was derived and is installed only if that version is still current.

If another writer has already advanced the same partition, publication fails rather than overwriting newer state. \thesystem{} invalidates the conflicting partition and allows the base-table transaction to commit normally. Subsequent queries that require that partition execute through the base-table fallback path until a later write successfully updates or rebuilds the state. If the transaction aborts, all accumulated mutations and staged auxiliary state are discarded.

% This protocol keeps auxiliary maintenance off the critical correctness path of the base transaction: maintenance conflicts can reduce performance by forcing fallback execution, but they do not change the outcome of the underlying database commit.

\subsection{Correctness Assumptions}\label{sec:correctness}
% - accelerator must be semantically equivalent to reference sql
% - uncomitted auxiliary updates invisible to other transactions
% - publication is atomic/versioned; readers never see partially updated state

\thesystem{} requires every generated accelerator to be semantically equivalent to its reference SQL query. This requirement applies to both stateless accelerators and stateful accelerators that answer queries using auxiliary structures. During synthesis, accelerator outputs are compared against the underlying DBMS using differential tests, and the final audit expands this test suite with cases that were not visible during optimization. Candidates that fail these checks are rejected.

For stateful accelerators, correctness also requires that auxiliary-state visibility respect the transaction semantics of the underlying DBMS. Updates produced by an uncommitted transaction remain transaction-local and are never visible to other readers. If the transaction aborts, its buffered mutations and any staged auxiliary-state changes are discarded. Whenever the system cannot establish that an auxiliary structure is safe to use, it falls back to execution over those tables.

\section{Evaluation}
% more crisp questions
Our evaluation answers four questions:
\begin{itemize}[leftmargin=*]
  \item How much does \thesystem{} accelerate analytical queries? (Section \ref{sec:read-performance})
  \item How does its performance change as writes become more frequent? (Section \ref{sec:update-performance})
  \item Does the maintenance policy adapt effectively to different write patterns? (Section \ref{sec:update-performance})
  \item What kinds of optimizations does the agentic pipeline discover? (Section \ref{sec:generated-examples})
\end{itemize}

\subsection{Experimental Setup}
% \begin{itemize}[leftmargin=*]
%   \item Models used
%   \item Machines used
%   \item Datasets/workload description
%   \item Base database systems (DuckDB, maybe PostgreSQL)
%   \item Competitor agentic systems (BespokeOLAP, GenDB)
% \end{itemize}

We evaluate \thesystem{} on two query workloads: TPC-H at SF=20~\cite{tpch} and SEC-EDGAR at SF=3. TPC-H provides a standard analytical workload with well-understood query patterns and optimization opportunities. We additionally use SEC-EDGAR, following GenDB \cite{gendb}, to evaluate whether the synthesis pipeline generalizes beyond workloads that are likely to be represented in LLM training data. SEC-EDGAR is constructed from real-world SEC financial-statement data from 2022-2024. GenDB~\cite{gendb} generates 1000 SQLSmith queries over this dataset, which we use for \thesystem{} evaluation.

Experiments run on a dual-socket server with two Intel Xeon Gold 6230 CPUs (40 physical cores, 80 hardware threads total), 376 GiB of memory, and a 1 TB NVMe SSD. Unless otherwise noted, query experiments use a single thread. 

To illustrate the generalizability of \thesystem{}, we evaluate on two DBMS backends: DuckDB and PostgreSQL 18.4. We compare against these base systems, as well as against GenDB \cite{gendb} and BespokeOLAP \cite{bespoke-olap}, and evaluate both stateless and stateful variants of \thesystem{}.

The synthesis pipeline uses GPT-5.6 Terra Pro for the analysis stage and DeepSeek V4 Pro for implementation, optimization, debugging, testing, and audit. Analysis runs through the OpenAI Agents SDK, while the code-oriented stages use the Codex \texttt{exec} harness.

\sparagraph{Synthesis cost.}
Synthesizing the DuckDB and PostgreSQL backends cost \$4.15 and \$3.15, respectively. Generating all 22 TPC-H stateful accelerators cost \$41.63, while the selected SEC-EDGAR stateful accelerators cost \$10.92. The 22 stateless TPC-H accelerators cost \$15.22.

\subsection{Steady-State Read Performance}
\label{sec:read-performance}

\begin{figure*}[t]
    \centering
    \includegraphics[width=\textwidth]{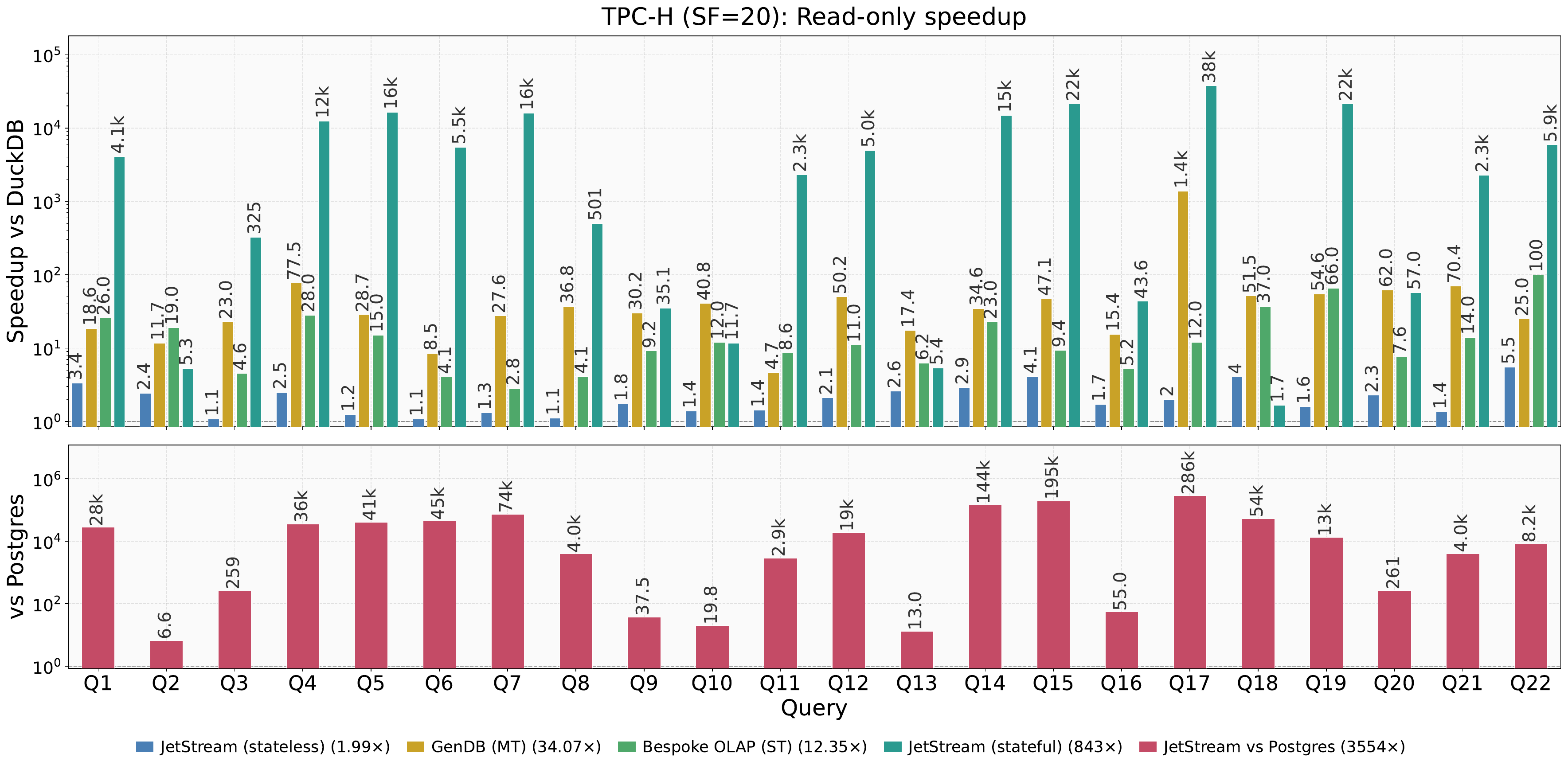}
    \caption{\textbf{\thesystem{} achieves large read-only speedups across TPC-H.}
    Per-query speedup at SF=20 relative to DuckDB and PostgreSQL.
    Stateful \thesystem{} achieves an $833\times$ geomean speedup over DuckDB and $3544\times$ speedup over PostgreSQL,
    substantially exceeding multi-threaded GenDB ($34.07\times$) and Bespoke OLAP
    ($12.35\times$), while stateless \thesystem{} achieves $1.99\times$.}
    \label{fig:tpch-sweep}
\end{figure*}

We first evaluate the steady-state read performance of generated accelerators on TPC-H and SEC-EDGAR. Figure~\ref{fig:tpch-sweep} reports per-query speedup on all 22 TPC-H queries at SF=20. Stateless \thesystem{} accelerators, which execute directly over the host DBMS without persistent auxiliary state, achieve a $1.99\times$ geomean speedup over single-threaded DuckDB. Allowing the pipeline to synthesize query-specific auxiliary state increases the geomean speedup to $833\times$, compared with $34.07\times$ for GenDB and $12.35\times$ for Bespoke OLAP. 
% Relative to PostgreSQL, the same stateful accelerators achieve a $3554\times$ geomean speedup.

The largest gains come from eliminating repeated analytical work rather than only making that work faster. A conventional TPC-H execution may scan and join large portions of the SF=20 database before producing a small aggregate result. Stateful accelerators generated \thesystem{} instead stores query-specific, parameter-independent sufficient statistics, such as per-date aggregates, compact multidimensional summaries, and range trees. On a hit, the accelerator would only have to read a few state partitions and performs a short final computation. For many queries, this reduces a multi-second scan, join, and aggregation pipeline to a sub-millisecond or few-millisecond lookup-and-compose path.

This more aggressive specialization also explains much of the gap with GenDB and Bespoke OLAP. Those systems generate workload-specific storage first, and then focus on speeding up query execution by specializing operator implementations. \thesystem{} goes further by additionally synthesizing persistent query-specific state that removes entire scans, joins, and aggregates from the read path. Additionally, the state that \thesystem{} generates is organized around the query attributes that vary across parameter bindings. For example, TPC-H Q14 is parameterized by ship date, so \thesystem{} maintains per-month revenue statistics and answers a bound query by reading the corresponding month and computing the final ratio. Range-oriented queries, for example, use trees or multidimensional summaries keyed by the attributes that define those ranges. The pipeline searches over these state organizations directly, allowing different queries to use different structures rather than conforming to a fixed, predetermined physical design, as GenDB and Bespoke OLAP do.

\begin{figure}[t]
    \centering
    \includegraphics[width=\columnwidth]{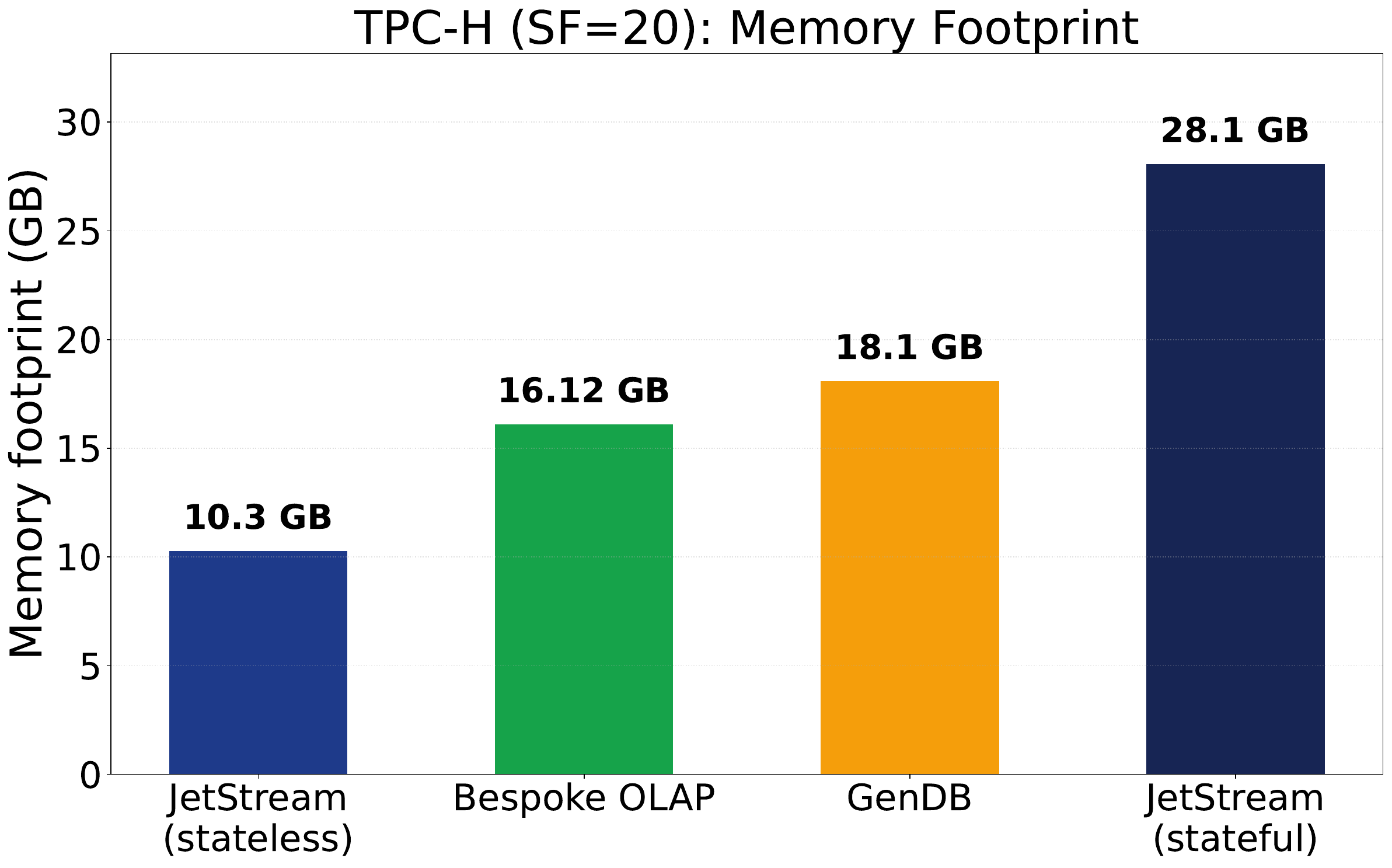}
    \caption{\textbf{Memory footprint on TPC-H at SF=20.}
    Stateful \thesystem{} uses additional memory to maintain query-specific
    auxiliary state.}
    \label{fig:memory-footprint}
    \vspace{-3mm}
\end{figure}

Naturally, these gains come with a space cost. As shown in Figure~\ref{fig:memory-footprint}, stateful \thesystem{} uses more memory than the stateless configuration and the generated-system baselines. This reflects an explicit time-space tradeoff: \thesystem{} uses additional query-specific auxiliary state to move more computation off the read path. The speedups in Figure \ref{fig:tpch-sweep} measure warm reads with valid auxiliary state and exclude the one-time cost of constructing that state. We account for the cost of keeping the state current under writes in Section~\ref{sec:update-performance}.

\begin{figure}[t]
    \centering
    \includegraphics[width=\columnwidth]{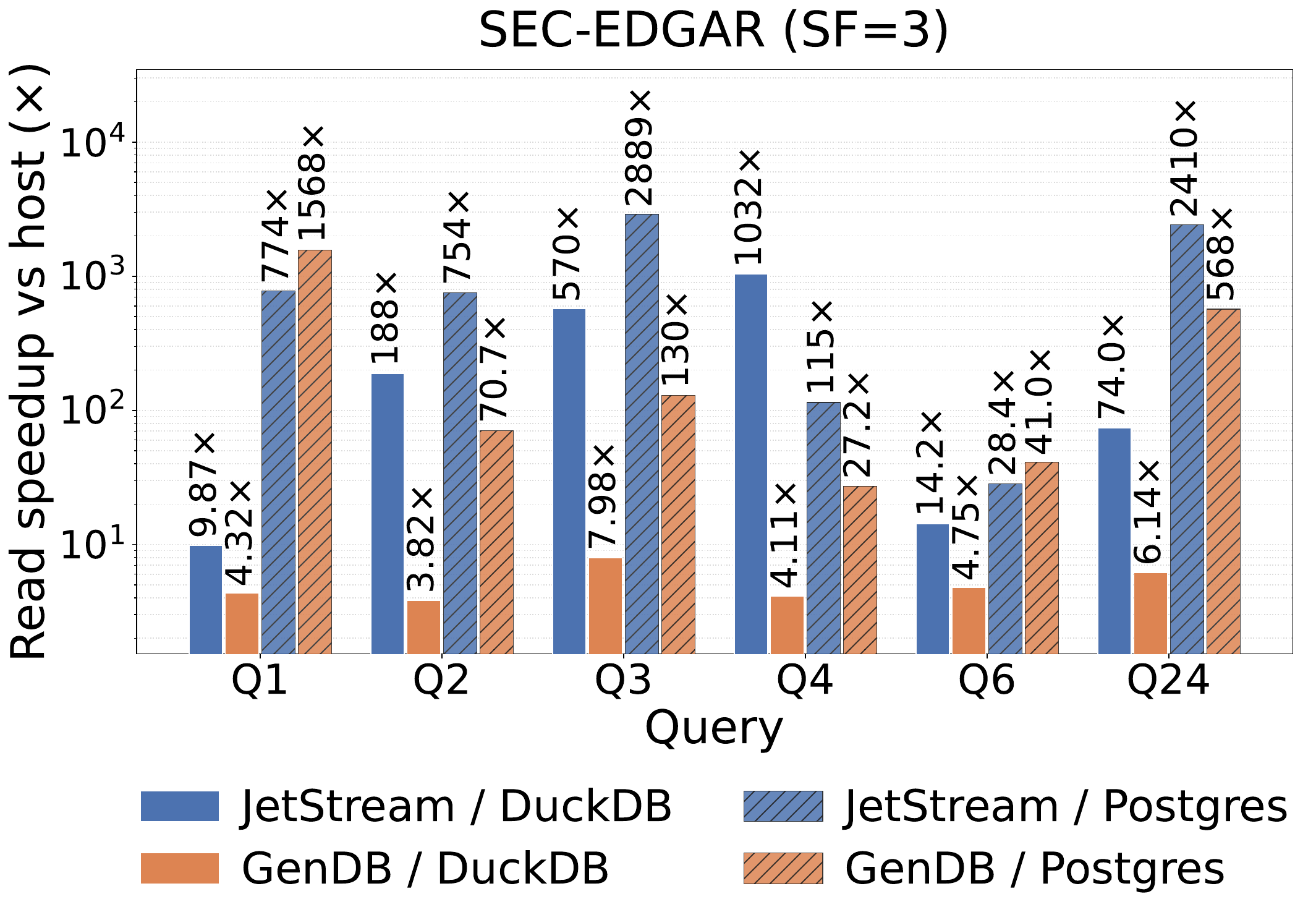}
    \caption{\textbf{\thesystem{} performance on unseen SEC-EDGAR queries.\thesystem{} achieves a $102\times$ geomean read-only speedup over DuckDB and $486\times$ over PostgreSQL. }}
    \label{fig:sec-edgar-speedup}
    \vspace{-3mm}
\end{figure}

Figure~\ref{fig:sec-edgar-speedup} shows that the gains are not specific to TPC-H. On SEC-EDGAR, \thesystem{} outperforms GenDB on most evaluated queries against both DuckDB and PostgreSQL. Since SEC-EDGAR was chosen to reduce the chance that the workload itself is well represented in LLM training data, the result therefore provides evidence that the pipeline can adapt to unfamiliar schemas, storage layouts, and query structures rather than relying only on well-known TPC-H optimizations. For example, SEC-EDGAR Q1 leads the pipeline to exploit the observed dictionary-over-flat representation of string columns; we discuss this case in Section~\ref{sec:generated-examples}. The gains also persist under updates: with point writes every 0.5\,s, \thesystem{} achieves an $8.8\times$ geomean workload speedup over DuckDB and $4.0\times$ over PostgreSQL, including both query and maintenance time. 

% \begin{itemize}[leftmargin=*]
%   \item Show end-to-end speedups of our system on TPC-H and another workload (SEC-EDGAR or TPC-DS or others)
%   \begin{itemize}
%     \item Workload should include writes + ad-hoc (unseen) queries
%     \item Compare the three versions of our system: native storage only, + auxiliary state, on open storage
%     \item Baseline: Base database system (DuckDB, PostgreSQL if time permits)
%   \end{itemize}
  
%   \item We need to compare with GenDB + BespokeOLAP, but should consider the
%   right setting. Maybe if the workload is write-heavy enough, we include it in
%   the headline figure.
% \end{itemize}
\subsection{Performance Under Updates}
\label{sec:update-performance}

We next evaluate how well \thesystem{} preserves its read advantage as the underlying database changes. We consider two complementary update regimes. First, we use the standard TPC-H refresh functions RF1 and RF2, which insert and delete batches corresponding to approximately 0.1\% of the database. These are comparatively large updates and stress the cost of rebuilding or incrementally maintaining substantial portions of auxiliary state. Second, we use synthetic point writes to isolate the behavior of fine-grained maintenance under much smaller but more frequent updates.

\begin{figure*}[t] 
\centering 
\includegraphics[width=0.8\textwidth]{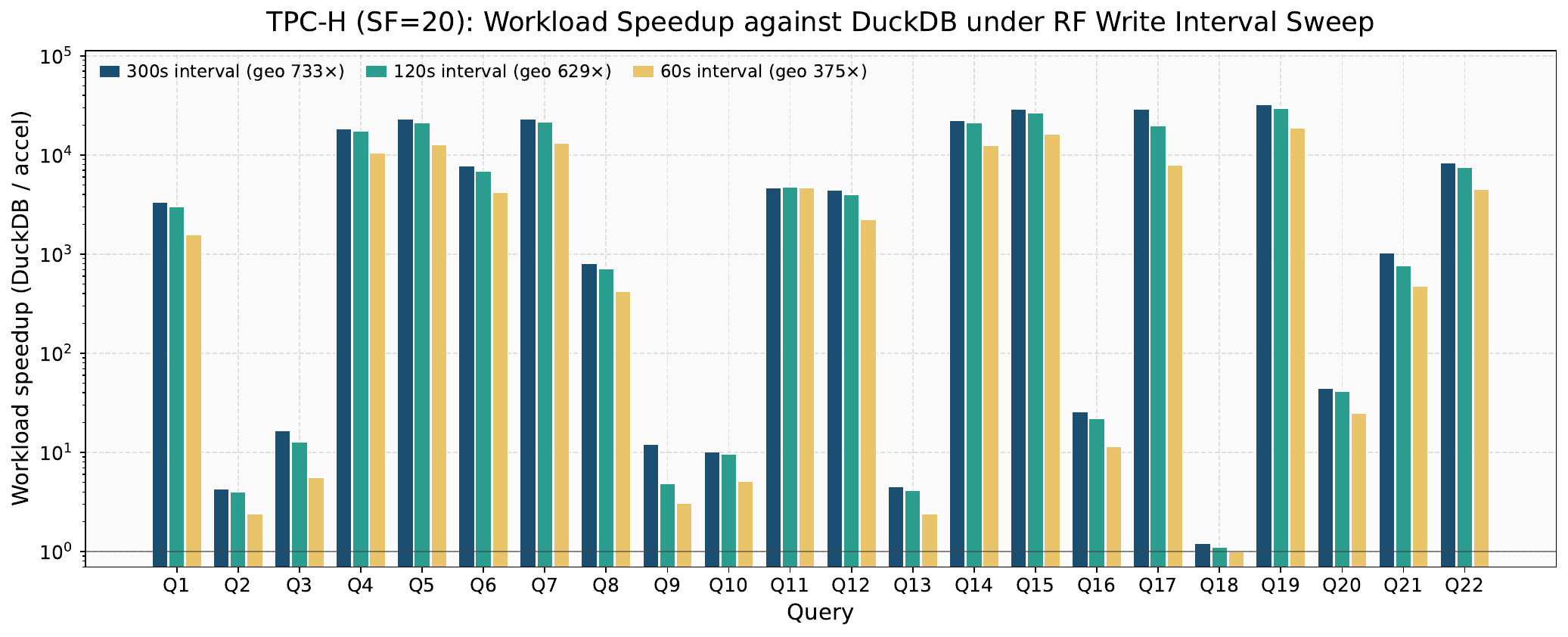} 
\caption{\textbf{\thesystem{} retains large speedups under TPC-H refreshes.} Per-query workload speedup over DuckDB at SF=20 for RF1/RF2 refresh intervals of 300\,s, 120\,s, and 60\,s. Geomean speedups: $733\times$, $629\times$, and $375\times$, respectively.} 
\label{fig:tpch-write-sweep} 
\end{figure*}

Figure~\ref{fig:tpch-write-sweep} shows the TPC-H refresh experiment. Even when a full RF update is issued once per minute, \thesystem{} retains a $375\times$ geomean workload speedup over DuckDB; at 120- and 300-second intervals, the geomeans rise to $629\times$ and $733\times$. The decrease with shorter intervals reflects the larger fraction of time spent maintaining or repairing auxiliary state, but the read savings still dominate for most queries.

This result also highlights the difference from read-only generated systems. GenDB and Bespoke OLAP can specialize aggressively when their generated state remains static, but do not provide a general mechanism for keeping that state consistent under updates. In contrast, \thesystem{} can incrementally update, rebuild, or invalidate and lazily repair state depending on the cost of the current write. As a result, the system can preserve much of its steady-state read benefit even when the base database is actively changing.

\begin{figure}[t]
    \centering
    \includegraphics[width=\columnwidth]{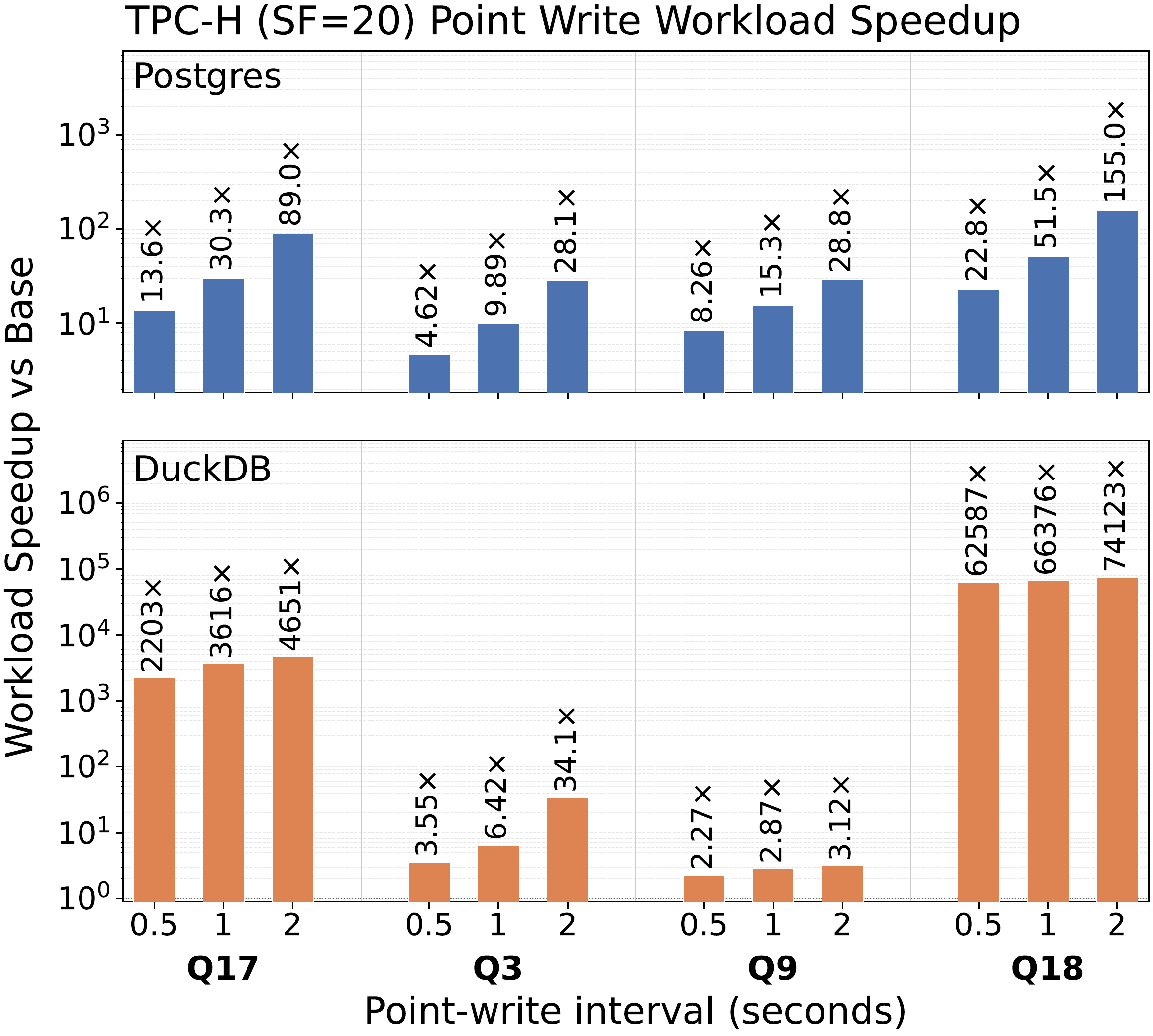}
    \caption{Per-query speedup over DuckDB and Postgres for \thesystem{} under synthetic point-write workload.}
    \label{fig:tpch-point-write}
    \vspace{-4mm}
\end{figure}

The point-write experiment in Figure~\ref{fig:tpch-point-write} examines the opposite regime: small updates arriving much more frequently. For queries whose auxiliary state admits cheap incremental maintenance, \thesystem{} continues to provide large speedups even at sub-second write intervals. Unlike the RF workload, these updates typically touch only a small number of rows and state partitions, allowing the maintenance policy to avoid expensive rebuilds.

\begin{figure}
    \centering
    \includegraphics[width=\columnwidth]{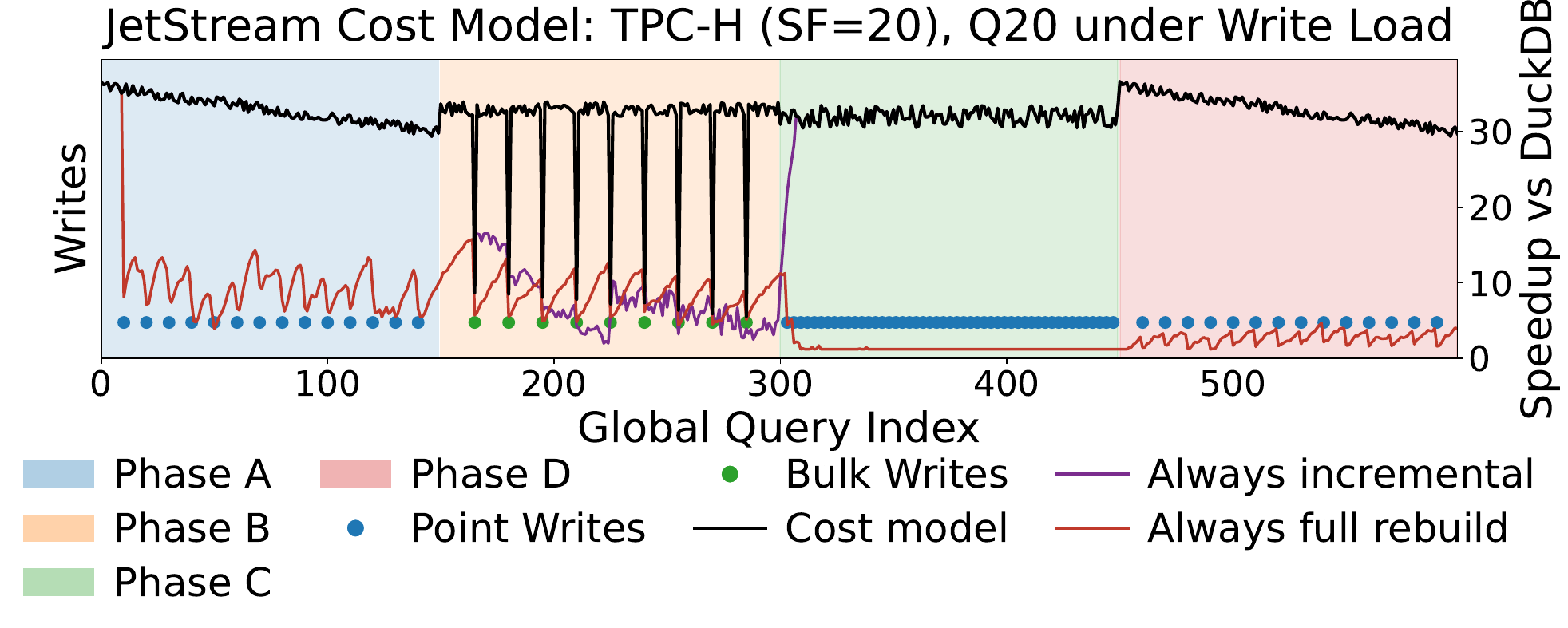}
    \caption{\thesystem{} adapts maintenance strategy to workload. Phase A is read-heavy with periodic point-writes; Phase B is read-heavy with periodic bulk writes; Phase C is write-heavy with frequent point-writes; Phase D returns to a read-heavy workload with periodic point writes.}
    \label{fig:cost-model-phases}
    \vspace{-4mm}
\end{figure}

Figure~\ref{fig:cost-model-phases} illustrates why the two update regimes---fine-grained point writes and larger bulk writes (such as those in the TPC-H refresh workload)---favor different maintenance strategies. Phase A is read-heavy with periodic point writes, Phase B remains read-heavy but switches to periodic bulk writes, Phase C increases the frequency of point writes, and Phase D returns to the read-heavy point-write regime.

During the point-write phases, only a small amount of auxiliary state is affected, so the selector favors incremental maintenance and closely tracks the always-incremental policy. An always-full-rebuild policy performs poorly in this regime because each small update incurs the cost of reconstructing much more state; when writes arrive frequently, rebuild work dominates the workload.

The tradeoff reverses under bulk writes. Applying many individual deltas becomes expensive, causing the always-incremental policy to lose much of its advantage. \thesystem{} instead switches to lazy rebuilding, after which accelerated reads resume. When the workload returns to point writes, the selector switches back to
incremental maintenance. Thus, neither fixed strategy performs well across both update regimes: \thesystem{} adapts its maintenance action to the size and shape of the current write workload, as described in Section~\ref{sec:maintenance}.
% \todo{duckdb postgres full refresh set speedup (mostly done)}

% \todo{running round robin experiment plot}
% \begin{itemize}[leftmargin=*]
%   \item Vary write frequency on x-axis, plot speedup on y-axis
%   (TPCH Refresh, best (round robin) and worst (invalidate all) case)
% \end{itemize}

\subsection{Generated Accelerator Examples}
\label{sec:generated-examples}
We next highlight three representative optimizations selected by \thesystem{}. These examples span both kernel optimization and auxiliary-state synthesis, and illustrate how the pipeline combines workload structure, profiling evidence, and read/write tradeoffs to arrive at non-obvious implementations.

Before discussing the generated workloads, we briefly summarize the relevant data in the two workloads. SEC-EDGAR is derived from public company filings. The \texttt{pre} table contains filing-line records; \texttt{adsh} identifies a filing, \texttt{stmt} is the financial-statement category, \texttt{rfile} is the report file category, and \texttt{line} gives the line number within the filing data. TPC-H models a sales database. The \texttt{customer} table stores customers, \texttt{orders} stores orders placed by those customers, and \texttt{lineitem} contains the individual items within each order. \texttt{part} stores product attributes such as brand, size, and container type. The examples below show how \thesystem{} reorganizes information from these base tables into query-specific state that reduces the work at read time.

\sparagraph{SEC-EDGAR Q1: layout-aware distinct aggregation.}
SEC-EDGAR Q1 summarizes filing-line records by statement and report-file category. For each group, it computes the number of records, the number of distinct filings, and the average line number. Evaluating it normally would require scanning roughly 9.6M qualifying rows, grouping on two low-cardinality string attributes, and tracking distinct values of the high-cardinality filing identifier \texttt{adsh}. Profiling showed that 99.7\% of input chunks used dictionary-encoded group keys with flat payload columns, while only about 379K distinct filing identifiers appeared across the 9.6M rows.

Based on this layout, \thesystem{} maintains persistent query-specific aggregation state. It maps each \texttt{adsh} once into a dense integer ID, and, for each statement and report-file group, stores the running row count, line-number sum, and compact distinct-filling representation (a bitmap) over those dense IDs. Because the observed input is overwhelmingly dictionary-encoded on the grouping columns, the accelerator can operate largely on compact IDs instead of repeatedly hashing and comparing strings. At read time, the result is produced directly from the maintained group state, by just looking at the bitmap cardinality and computing the average line number from the running count.

The optimization emerged through several failed attempts. Earlier variants introduced a custom string hash set or fast paths for layouts that profiling later showed were uncommon; both failed to improve the dominant cost and in some cases regressed performance. Phase timers and layout counters instead pointed to repeated distinct-string processing on the mixed dictionary-over-flat path, which led to the retained map-and-bitmap design.

\sparagraph{TPC-H Q13: q-gram postings with sparse histogram updates.}
TPC-H Q13 asks how many customers have each possible number of qualifying orders. For a bound comment pattern, it excludes orders whose comments match that pattern, left-joins the remaining orders to customers, counts the qualifying orders per customer, and then constructs a histogram over those counts. The expensive work is therefore both parameter-dependent and spread across several operators: comment filtering, the customer--order join, and two levels of aggregation.

\thesystem{} separates the parameter-independent work from the parameter-dependent correction. Its persistent auxiliary state contains a q-gram inverted index over order comments, a mapping from orders to customers, each customer's baseline order count, and a baseline histogram of customer order counts. At read time, the bound comment pattern is decomposed into q-grams, whose posting lists identify candidate matching orders. \thesystem{} verifies the exact \texttt{LIKE} predicate only on those candidates, maps the matched orders to affected customers, adjusts only those customers' counts, and applies sparse corrections to the baseline histogram. This avoids rescanning all orders and recomputing counts for every customer.

The analysis stage considered simpler alternatives, including a compact customer-comment intermediate and an inverted index followed by a full scan over customers. These designs removed part of the original work but left a scan-sized residual cost. The selected design combines selective search with sparse histogram correction, removing both the orders scan and the all-customer scan for selective bindings. It achieved roughly $8.6\times$ best-valid speedup while retaining an advantage under a 10\% write mix.

\sparagraph{TPC-H Q19: write-aware multidimensional range aggregation.}
TPC-H Q19 computes total discounted revenue for three parametrized regions of the product space. Each branch constrains brand, container family, \texttt{part} size, and \texttt{lineitem} quantity, together with fixed shipping predicates. A normal execution scans \texttt{lineitem}, joins it with \texttt{part}, applies the three predicate branches, and sums the surviving revenue even though the final result is a single scalar.

\thesystem{} instead maintains a persistent join-derived revenue summary keyed by brand and container family, with each entry organized as a two-dimensional range structure over \texttt{part} size and \texttt{lineitem} quantity. The stored values represent summed discounted revenue from the corresponding joined \texttt{lineitem}-\texttt{part} region. At read time, each of the three SQL branches becomes one range-sum lookup, and the final answer is the sum of those three probes. This replaces the fact-table scan, join, and most of the predicate evaluation with a small number of logarithmic-time range queries.

The analysis stage also considered a two-dimensional prefix-sum representation, which would make reads even cheaper. It rejected that design because a point update would modify an unbounded suffix of the prefix structure, creating a large maintenance cost at higher scale factors. The selected range tree instead keeps both reads and point updates logarithmic, illustrating how \thesystem{} trades a small amount of additional read work for substantially cheaper maintenance. The resulting accelerator achieved a best-valid dual-axis score of roughly $2638\times$ and remained hundreds of times faster under a 10\% write mix.

\section{Related Work}\label{sec:related-work}

\sparagraph{Synthesized accelerators.}
Bespoke OLAP~\cite{bespoke-olap} and GenDB~\cite{gendb} both propose techniques
to generate specialized query execution code by leveraging LLMs. However, as
discussed in \Cref{sec:intro}, these systems have practical limitations.
GenDB rewrites the underlying storage and generates query executors for
pre-defined templates~\cite{gendb}. As a result, supporting writes requires
re-running its storage optimization, and GenDB is unable to run unseen (e.g.,
ad-hoc) queries.
Bespoke OLAP is similar, but it retains a fallback database system for unseen
queries. This fallback effectively requires two copies of the data: one for
accelerated queries, and one in the database system. Like GenDB, it also cannot
support writes while preserving accelerated performance.

In contrast, \thesystem{} integrates its accelerators into a full featured
database system and relies on the database's native storage. This design is
practical: it requires only one copy of the data, queries for which an
accelerator is available can use it directly, and unseen queries simply run
unchanged on the underlying database.
\thesystem{} also leverages auxiliary state to accelerate specific queries, but
it generates incremental maintenance logic for this state, allowing it to
preserve most of its accelerated performance while delivering fresh results in
the presence of writes.

GPT-DB~\cite{gptdb-trummer23} also generates code to execute a query, but does
so interactively.
Castor~\cite{castor-feser20} is a classical approach to specializing the data
layout to a static templated query workload; it does not support writes to the
underlying data.

\sparagraph{Database extensibility.}
Database extensions~\cite{dbextsurvey-kim25, pgextensions, duckdb-extensions},
user defined functions~\cite{udfprism-arch25, udo-sichert22}, and declarative
sub-operators~\cite{dsos-jungmair23} all provide mechanisms for adding
functionality to a database system, which can include accelerators.
However, with these mechanisms, end users are responsible for identifying and
implementing the acceleration techniques, maintaining any auxiliary state,
rewriting their queries to use the accelerators when beneficial, and porting
their accelerator implementations to all the database systems they want to
support.
Recent work on Tailwind~\cite{tailwind-yu26} addresses most of these limitations
by providing a framework for non-invasively integrating accelerators into
database systems that support data import/export. Users define and implement
parameterized accelerator families, and Tailwind then configures and selects
accelerator instances for a given workload.
\thesystem{} goes further by using agents to automatically identify and
implement acceleration opportunities and generate auxiliary-state maintenance
logic. When possible, it uses UDFs and database extensions to integrate the
resulting accelerators with the underlying database system.

\sparagraph{Learned specialization.}
There is a long line of prior work that uses machine learning techniques to
specialize a database system or data infrastructure to specific query
workloads~\cite{copyright-sudhir21, tsunami-ding20, flood-nathan20,
qdtree-yang20, sagedb-ding22, brad-yu24, neo-marcus19, bao-marcus21,
snarf-vaidya22, lsort-kristo20, plbf-vaidya21, treeline-yu22,
selfdrivingoperation-pavlo21}.
At a high level, these techniques define a parameterized space of possible
system configurations and leverage learned models or analytical models to
predict performance across this space, allowing the system parameters to be
optimized for a given workload.
\thesystem{} is similar in spirit in that it specializes query execution to a
given workload, but it does not require a manually defined parameterized
configuration space; instead, LLM agents perform the workload analysis, code
generation, and optimization.

\sparagraph{Query compilation.}
Database systems such as Umbra~\cite{umbra-neumann20},
HyPer~\cite{hyper-kemper11}, Redshift~\cite{redshiftindustry-armenatzoglou22},
etc., also generate code to implement their chosen physical query
plans~\cite{querycompilation-neumann11}.
However, classical query compilation is designed for generality: it uses a fixed
code generation recipe for each operator in query
plan~\cite{querycompilation-neumann11}.
In contrast, \thesystem{} generates custom code that is specialized to a
specific query and dataset, allowing it to employ query and data-specific tricks
(e.g., bitpacking) and auxiliary state to accelerate the query.

\sparagraph{Incremental view maintenance.}
\thesystem{} incrementally maintains any auxiliary state it chooses to generate,
which is similar to incremental view maintenance~\cite{ivm-zhou07, dbsp-budiu23,
dbtoaster-ahmad09, noria-gjengset18, naiad-murray13}. However, the key
difference is that incremental view maintenance comprises algorithms for
maintaining state computed from a SQL expression. In contrast, \thesystem{}
generates incremental maintenance code for arbitrary auxiliary state. In other
words, the auxiliary state \thesystem{} uses is not necessarily always a
database view.

\iffalse
\begin{itemize}[leftmargin=*]
  \item \textbf{Synthesizing accelerators.} BespokeOLAP, GenDB, GPT-DB, Castor
  \item \textbf{Database extensibility.} UDFs, UDOs, declarative sub-operators (DSOs), Tailwind (user provided accelerators)
  \item \textbf{Learned system specialization.} SageDB, Bao/Neo, Tsunami/Flood/Qdtree, TreeLine, BRAD
  \item \textbf{Query compilation.} Neumann, HyPer/Umbra
  \item \textbf{Incremental view maintenance.} DBSP, materialize, noria, DBToaster, other classical work
  % \item \textbf{Coding agents?}
\end{itemize}
\fi

\section{Conclusion}

We presented \thesystem{}, a framework for generating query-specific database accelerators that extend an existing DBMS, rather than replace it. \thesystem{} combines a staged, measurement-driven synthesis workflow with a fixed engine-neutral substrate and generated backend adapters, allowing accelerators to execute directly over native database storage, and, when useful, maintain query-specific auxiliary state.

The resulting accelerators span a wide range of designs, from storage-aware stateless kernels to persistent indexes. Since \thesystem{} treats reads and writes jointly, it can choose among incremental maintenance, rebuilds, and lazy repair while preserving the transactional semantics of the underlying database. Across established datasets and unseen workloads, \thesystem{} is able to generate accelerators with large read speedups while retaining substantial gains under ongoing updates.

More broadly, our results suggest that LLM-based database specialization need not require synthesizing an entire replacement engine or assuming static data. Instead, a production DBMS can remain responsible for canonical storage, transactions, recovery, and general SQL, while generated components only specialize the known expensive parts of a workload. This separation offers a practical path towards increasingly aggressive, workload-specific optimization without giving up the functionality and reliability of mature database systems.

\bibliographystyle{ACM-Reference-Format}
\bibliography{codegen}

\end{document}